\documentclass[letter,twocolumn]{jpsj2}
\def\runtitle{
Interpocket polarization model
for magnetic structures
in rare-earth hexaborides
}
\def\runauthor{Yoshio {\sc Kuramoto} and Katsunori {\sc Kubo}}

\title
{
Interpocket polarization model for magnetic  structures
in rare-earth hexaborides
}

\author
{
Yoshio {\sc Kuramoto}\footnote{E-mail: kuramoto@cmpt.phys.tohoku.ac.jp} and Katsunori {\sc Kubo}\footnote{E-mail: katukubo@cmpt.phys.tohoku.ac.jp}
}

\inst
{
Department of Physics, Tohoku University, Sendai 980-8578
}

\recdate
{
\hspace{4cm}
}

\abst
{
The origin of peculiar magnetic structures
in cubic rare-earth (R) hexaborides RB$_6$
is traced back to
their characteristic band structure.
The three sphere-like Fermi surfaces induce
interpocket polarization of the conduction band
as a part of a RKKY-type interaction.
It is shown for the free-electron-like model
that the interpocket polarization gives rise to
a broad maximum in the intersite interaction $I(\mib q)$
around $\mib q=(1/4,1/4,1/2)$ in the Brillouin zone.
This maximum is consistent with the
superstructure observed in R=Ce, Gd and Dy.
The wave-number dependence of  $I(\mib q)$ is independently extracted from
analysis of the spin-wave spectrum measured for NdB$_6$.
It is found that $I(\mib q)$ obtained from fitting the data has a similarly to that derived by the interpocket polarization model, except that the absolute maximum now occurs at $(0,0,1/2)$ in consistency with the A-type structure.
The overall shape of $I(\mib q)$ gives a hint toward understanding an incommensurate structure in PrB$_6$ as well.
}

\kword
{
GdB$_6$, NdB$_6$, PrB$_6$, CeB$_6$, hexaborides, RKKY interaction, elastic constant, quadrupolar interaction
}

\begin{document}
\sloppy
\maketitle


Interest in cubic rare-earth (R) hexaborides RB$_6$ comes mainly from the rich structure in their ordered phases.  The best studied example is CeB$_6$ which undergoes a quadrupole (orbital) order at  3.3 K and then a magnetic order at 2.4 K \cite{effintin}.   The magnetic ground state is characterized by double-$\mib k$ structure with wave vectors $(1/4, \pm 1/4, 1/2 )$ in units of the reciprocal lattice parameter $2\pi/a$.  Since the orbital order is superimposed on the magnetic order, it has been
suspected that the orbital degeneracy in the crystalline electric field (CEF) ground state $\Gamma_8$ plays an important role \cite{kusu}.
Recently, however, neutron scattering experiment on GdB$_6$ has detected an equivalent wave vector in the ordered phase below 15 K \cite{kuwahara}.   Since the trivalent Gd ion has a half-filled 4f shell without orbital degrees of freedom, the order at $\mib k = (1/4, 1/4, 1/2 )$ should have an origin which does not depend so much on the particular configuration of 4f electrons.  It is known that PrB$_6$ also has the same wave number in the magnetically ordered ground state below 4.2 K, but the
intermediate phase between 4.2 K and 6.9 K has an incommensurate structure \cite{burlet}.
On the other hand,  the ground state of NdB$_6$ has a simple antiferromagnetic structure called the type I (or A-type) with alternating plane polarized along and against (0,0,1) \cite{mccarthy,erkelens}.

In this paper we propose a simple model to understand the origin of these structures from a unified point of view.
The basic observation is that the Fermi surface of RB$_6$ consists of three nearly spherical pieces centered on the $X$ points $X_x=(1/2,0,0), X_y=(0,1/2,0), X_z=(0,0,1/2)$ in the Brillouin zone. The RKKY interaction involves interpocket polarization, which has a new characteristic wave vector $\mib K_3 = (1/2,1/2,0)$ which connects $X_x$ and $X_y$, and equivalent ones.
Just like the ordinary RKKY interaction can bring about the antiferromagnetic ordering by halving the reciprocal lattice vector, the halving of the characteristic wave vector $(1/2,1/2,0)$ can bring about the ordering at
$(1/4, 1/4, 1/2)$.

Let us consider the case of GdB$_6$ where the 4f electrons have only the spin degrees of freedom.  The exchange interaction
between a 4f-spin $\mib S_i$ at $\mib R_i$ and conduction electrons is taken to be
\begin{equation}
H_{df}  = \frac{J}{N}\sum_{\mib k,\mib p}\sum_i
W_{\mib k,\mib p}
{\rm e}^{i (\mib{p-k})\cdot\mib R_i}
\mib S_i\cdot\mib\sigma_{\alpha\beta}
c_{\mib k\alpha}^\dagger c_{\mib p\beta},
\label{H_df}
\end{equation}
where $N$ is the number of lattice sites, and
$JW_{\mib k,\mib p}$
is determined by the exchange integral involving 4f wave functions and
the conduction states.
We follow the previous argument \cite{schmitt} to derive $JW_{\mib k,\mib p}$.
In analogy with the APW method we consider
a muffin-tin sphere centered at the origin.
The Bloch function $\psi_{\mib k} (\mib r)$ of the conduction band is expanded inside the sphere as
\begin{equation}
\psi_{\mib k} (\mib r) = \frac{1}{\sqrt N}\sum_\lambda R_{k\lambda} (r)\sum_{\Gamma\gamma}
d_{\Gamma\gamma}^{(\lambda)}(\mib k) Y_{\Gamma\gamma}^{(\lambda)} (\hat r),
\label{decomposition}
\end{equation}
where $R_{k\lambda} (r)$ describes the radial part with
orbital index $\lambda$,
and  $Y_{\Gamma\gamma}^{(\lambda)} (\hat r)$ with $\hat r =\mib r/r$ is the cubic harmonics for
the point-group representation $\Gamma$ and its component $\gamma$.
We neglect the $k$-dependence of $R_{k\lambda} (r)$
since the extent of 4f electrons is smaller than that of 5d electrons which contribute
dominantly to the exchange.
Because the orbital angular momentum is zero in Gd$^{3+}$, the exchange integral  becomes diagonal with respect to the azimuthal quantum number of 4f states, and to $(\Gamma,\gamma)$.
Thus the exchange interaction becomes isotropic with a factor
\begin{equation}
W_{\mib k,\mib p} = \sum_{\Gamma\gamma}
d_{\Gamma\gamma}^{(5d)}(\mib k)^*
d_{\Gamma\gamma}^{(5d)}(\mib p).
\label{weight}
\end{equation}

The RKKY interaction $I(\mib q)$ is  given by
$$
I(\mib q) = \frac{2J^2}{N} \sum_{\mib k} |W_{\mib k,\mib{k+q}}|^2 \frac{
f(\epsilon_{\mib {k+q}})-
f(\epsilon_{\mib k}) }
{\epsilon_{\mib k}- \epsilon_{\mib {k+q}}}
$$
with $f(\epsilon)$ being the Fermi function.
The intrapocket contribution to $I(\mib q)$ comes from such $\mib k$ and $\mib{k+q}$ that belong to the same pocket of the conduction band.
In addition, there arises the interpocket contribution explained earlier.
Let us take the free-electron-like dispersion and set
$|W_{\mib k,\mib{k+q}}|^2$ constant in order
to see the consequence of the interpocket contribution in the simplest manner.
In order to keep the lattice periodicity,
we take summation over the reciprocal lattice vectors $\mib G$
rather than restricting the $\mib k$-summation within the Brillouin zone.
The $\mib G$-summation corresponds to inclusion of higher energy bands.
Namely we introduce
$$
\tilde\Pi (\mib q) = \sum_{\mib G}F(\mib q+\mib G) \Pi (\mib q+\mib G),
$$
where $F(\mib q+\mib G)$ is a form factor to be specified later, and
$\Pi (\mib q)$ is the Lindhard function multiplied by the partial density of states at the Fermi level.   The Fermi wave number $k_F$ is given by
$
k_F a/\pi =\pi ^{-1/3} = 0.9656/\sqrt 2,
$
which means that the three spherical Fermi surfaces barely touch with one another.
In the real RB$_6$ system, the Fermi surface also contains a fine structure along $(1,1,0)$ and equivalent directions \cite{onuki,harima}.

Adding both intrapocket and interpocket contributions we
obtain for $\mib q \in$ Brillouin zone:
$$
I(\mib q)
=J^2 [3\tilde\Pi (\mib q)
+ 2\sum_{i=1}^3 \tilde\Pi (\mib q-\mib K_i)]
\equiv J^2\chi(\mib q),
$$
where $3\tilde\Pi (\mib q)$
accounts for the three equivalent pockets, and
$\tilde\Pi (\mib q-\mib K_i)$
describes the interpocket polarization.
The $\mib K_i$'s are given by
$
\mib K_1 =(0,1/2,1/2),
\mib K_2 =(1/2,0,1/2),
\mib K_3 =(1/2,1/2,0)
$
in units of $2\pi/a$.
The factor 2 for the interpocket term enters because each pocket can be both starting and ending states of the transition.
For simplicity we take
the form factor such that $F(\mib k)=1$ if
$|k_\alpha| < 6\pi/a$ for all components $\alpha = x,y,z$ and
zero otherwise.
The choice of the cut-off in the form factor hardly influences the $\mib q$-dependence of the RKKY interaction, although it does influence the absolute value. Specifically with $F(\mib{q+G})=1$ for all $\mib G$, $\tilde\Pi (\mib q)$ diverges logarithmically by summation over $\mib G$.

\begin{figure}[t]
\includegraphics[width=8.5cm]{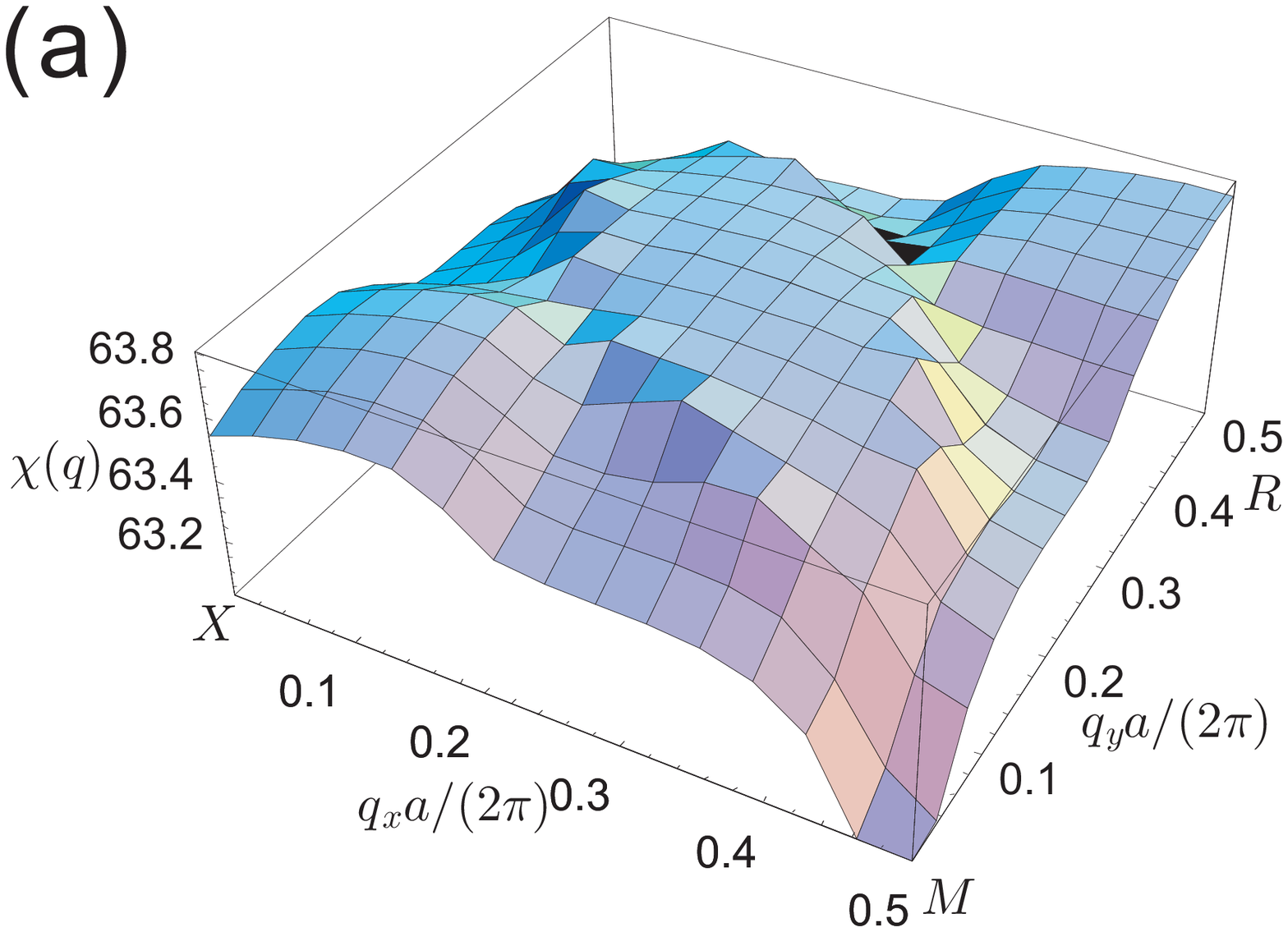}\\

\includegraphics[width=8cm]{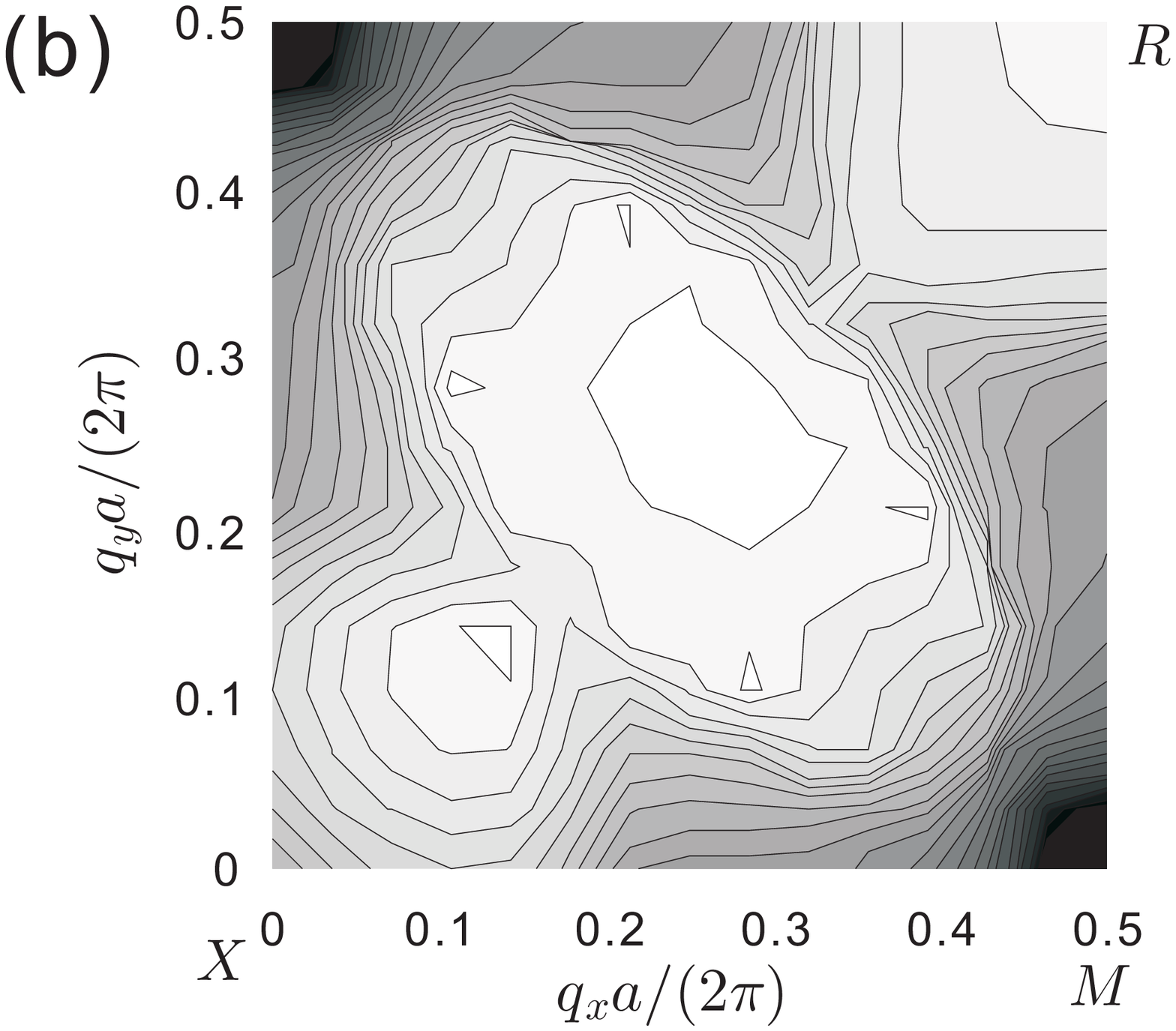}
\caption{
Wave-number dependence of the intersite interaction
in the {\it X-M-R} plane:  (a) three-dimensional plot; (b) contour plot.
}
\label{fig:I(q)}
\end{figure}

Figure \ref{fig:I(q)} shows $\chi(\mib q)$ in the
{\it X-M-R} plane of the Brillouin zone with
$X=(0,0,1/2), M=(1/2,0,1/2)$ and $R=(1/2,1/2,1/2)$.
The unit of ordinate is such that $\Pi(0)=1$,
and the large numerical value of $\chi(\mib q)$ comes from summation over $\mib G$.
For the intersite interaction, only the variation in the $\mib q$-space is relevant since the average of $\chi(\mib q)$ represents the intra-site contribution.

We have also made a scan of $\chi (\mib q)$ along $(1/4,1/4,q_z)$ and found that the peak indeed occurs at $q_z=1/2$.
It is apparent that the interaction favors the magnetic order near the center of the {\it X-M-R} plane, namely around $(1/4,1/4,1/2)$.
Since the ridge extends more toward $M$ rather than $R$, an incommensurate structure with $q_x \neq q_y$ can be realized
by slight change of the system parameters.

We now analyze in more detail the character of the conduction band,
which consists mainly of
$t_{2u}$ molecular orbitals of 2p electrons in B$_6$ clusters hybridized with
$e_g $ orbitals of 5d electrons.
One of the $t_{2u}$ orbitals has the angular dependence $z(x^2-y^2)$ if seen from the center of the B$_6$ cluster, and hybridizes best with the 5d $x^2-y^2$ orbital at neighboring rare-earth sites.  Since $z(x^2-y^2)$ changes sign below and above the B$_4$ plane, the wave number $(0,0,1/2)$ gains the bonding energy optimally \cite{harima}.  Thus the bottom of the conduction band goes to $X_z$ and equivalent points.
For $\Gamma =e_g$
we use a simplified notation
$ ( \gamma |\mib k) = d_{\Gamma\gamma}^{(5d)}(\mib k) $
with $\gamma$ being either the state $x^2-y^2$ or $3z^2-r^2$.
Then we have a large amplitude $( x^2-y^2 |X_z )$,
while $( 3z^2-r^2 |X_z )$ is negligible.
At another point $X_x=(1/2,0,0)$, the wave function has the character of $y^2-z^2$ which can also be represented by
\begin{equation}
|y^2-z^2\rangle = -\frac 12|x^2-y^2\rangle -\frac{\sqrt 3}{2} |3z^2-r^2\rangle,
\end{equation}
with use of the basis set at $X_z$.
Thus the orbital flip from $|x^2-y^2\rangle$ to $|y^2-z^2\rangle$ can take place even with the cubic symmetry.
We note that the finite overlap does not contradict with the orthogonality of Bloch functions with different $\mib k$.

The relative weight of the interpocket polarization against the intrapocket one should influence the detailed behavior of $I(\mib q)$.
We estimate from the above argument the weight factor $W_{\mib k,\mib p}$
for $\mib k =X_z$ and $\mib p =X_x$ relative to $W_{\mib k,\mib k}$ as
$$
W_{\mib k,\mib p} /W_{\mib k,\mib k}
\sim -1/2.
$$
These points $X_z$ and $X_x$, however, are not on the Fermi surface.
In the region where two pieces of the Fermi surface almost touch with each other, the interpocket contribution connecting the nearby $\mib k$ states should have a larger weight factor than the intrapocket one with remote $\mib k$ and $\mib p$.
We have made a tight-binding calculation taking
the $e_g$  and $t_{2u}$ orbitals,
and examined the character
of the wave functions at various points in the Brillouin zone.
It is found that 2p-electron weight is larger than the 5d-electron weight in general, and the latter changes gradually from $|x^2-y^2\rangle$ to $|y^2-z^2\rangle$ as $\mib k$ moves from $X_z$ to $X_x$.
In a future work, we shall evaluate $W_{\mib {k,p}}$ by using realistic wave functions.

The presence of orbital degeneracy in rare-earth ions other than Gd makes it necessary to consider more complicated form of $H_{df}$.
Namely not only the spin exchange interaction but multipole interactions also enter \cite{schmitt,kasuya,teitelbaum}.
As long as the conduction band consists purely of $e_g$ for the 5d electron part,
the wave-number dependence of the multipole intersite interactions is the same as that of the exchange interaction.
Actually, however, $t_{2g} (=\Gamma_5)$ also enters into eq.(\ref{decomposition}).
In the presence of orbital degeneracy,
$W_{\mib {k,p}}$ is no longer given by eq.(\ref{weight}) but with different weights for each $\Gamma$ \cite{schmitt}.
Moreover, hybridization between 4f electrons and boron 2p electrons may become important in the open-shell case.  The hybridization constitutes another mechanism of the intersite interaction.

With these complications in mind
we proceed to analysis of the exchange interaction in NdB$_6$ where the spin-wave spectrum has been measured.
The dipole part of $H_{df}$ can be taken in the same form as eq.(\ref{H_df}) except that $\mib S_i$ is replaced by the angular momentum operator $\mib J_i$ with $J=9/2$.
With only the magnetic intersite interaction, the easy axis of the magnetic moment should be along $(1,1,1)$ \cite{uimin}.
Actually the moment is parallel or antiparallel to $(0,0,1)$, which has been explained in terms of ferroquadrupolar interaction \cite{pofahl,sera}.
With inclusion of the magnetic and $\Gamma_3$-type quadrupolar interactions, we consider the following model:
$$
  H=-\sum_{(i,j)} I_{ij} \mib{J}_i \cdot \mib{J}_j
  -g_3' \sum_{i} \left(\langle O^0_{2}\rangle O^0_{2i}+3\langle O^2_{2}\rangle O^2_{2i}\right),
$$
where we assume that the average of the quadrupole moment does not depend on a site.
Other interactions such as the $\Gamma_5$-type quadrupolar interaction \cite{nakamura} are neglected since they do not affect the spin-wave spectrum.

The CEF ground state is $\Gamma_8^{(2)}$, which is four-fold degenerate, and the first excited state lies 132$\sim$135  K above \cite{lowenhaupt,pofahl}.
We introduce the Pauli matrix $\sigma^\alpha (\alpha =x,y,z)$ to describe the Kramers pair, and another Pauli matrix
$\tau^\alpha$ to describe the orbital pair in the $\Gamma_8^{(2)}$ quartet.
Then the angular momentum operator $J^\alpha $ within the $\Gamma_8^{(2)}$ subspace is written as
$$
J^\alpha = \frac 12 (\xi +\eta T^\alpha)\sigma^\alpha,
$$
where $\xi = -0.883$ and $\eta = -4.712$ are numerical constants
corresponding to the Lea-Leask-Wolf \cite{LLW} parameter $x=-0.82$
\cite{pofahl}.
The orbital effect on the magnetic moment is described by
$T^\alpha$ with $T^{z}=\tau^z$ and
$$
  T^{x,y}=-\tau^z/2\pm\sqrt{3}\tau^x/2.
$$

In the N\'{e}el state the $\Gamma_8^{(2)}$ quartet undergoes a Zeeman splitting by the molecular field.  This splitting induces a finite quadrupole moment which is enhanced by positive $g_3'$.
Then the lowest level is characterized by $(\tau^z, \sigma^z) = (+, \uparrow)$ in the A-sublayer and $(+, \downarrow)$ in the B-sublayer.
One may expect two branches corresponding to excitations
$$
(+, \sigma) \rightarrow (\pm, -\sigma),
$$
with intensities $I_\pm$.   Here $\sigma =\uparrow, \downarrow$ in the A- and B-sublayers, respectively.
The intensity ratio $I_+/I_- $ is given by
$$
I_+/I_-  = (2\xi/\eta -1)^2/3 \sim 0.13.
$$
Thus we identify the observed branch as the inter-orbital transition
$(+, \sigma) \rightarrow (-, -\sigma)$,
and assume that intra-orbital branch was not detected because of the small intensity.

By neglecting the small matrix elements for intra-orbital magnetic excitations, we obtain a reduced model
which keeps only the two levels leading to $I_-$.
Assuming the A-type antiferromagnetic structure with $\mib{Q}=(0,0,1)$, we obtain the excitation spectrum by
the standard spin-wave theory as
\begin{equation}
  \omega^2_{\mib{q}}=[J(\mib{q})-\Delta ][J(\mib{q}+\mib{Q})-\Delta ],
\label{omega^2}
\end{equation}
where $J(\mib{q})$ is the Fourier transform of
$J_{ij}=\left(3 \eta^2/16\right) I_{ij},$
and
\begin{equation}
\Delta=\frac{8\xi(\xi+\eta)}{3\eta^2} J(\mib{Q})+\frac 92 (\xi-\eta-9)^2g_3'.
\label{delta}
\end{equation}
The spectrum given by eq.(\ref{omega^2}) was also postulated
in the previous work, where $\omega_{\mib{q}}$ experimentally measured was fitted by intersite interactions up to third neighbors.  The authors of ref.\citen{erkelens} noted that the calculated N\'{e}el  temperature was about half of the experimental one, 8$\sim$9 K.
We point out further that the substantial softening around $(1/4,1/4,0)$ was not reproduced by the previous fit.
\begin{figure}
\includegraphics[width=8cm]{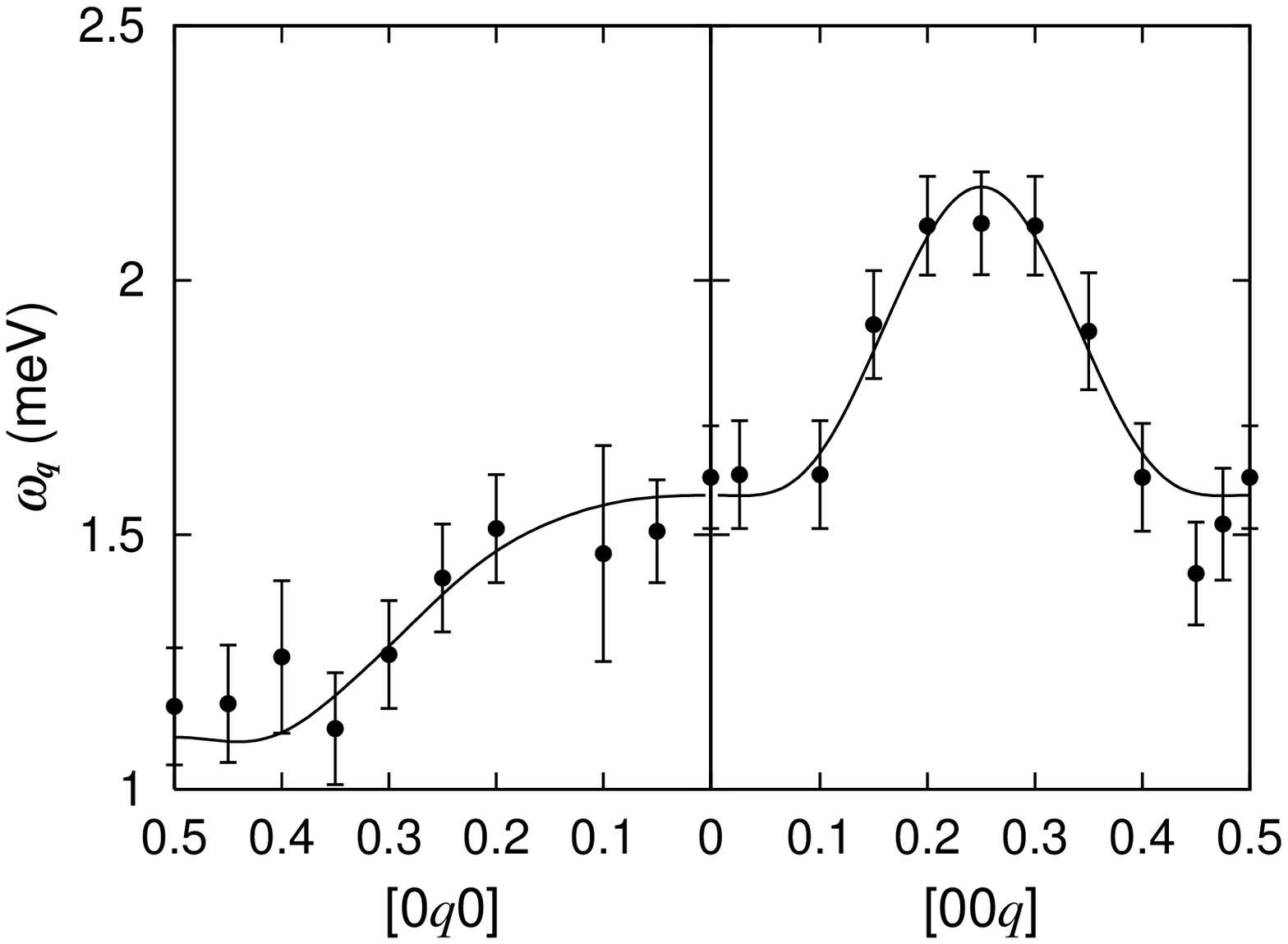}

\includegraphics[width=8cm]{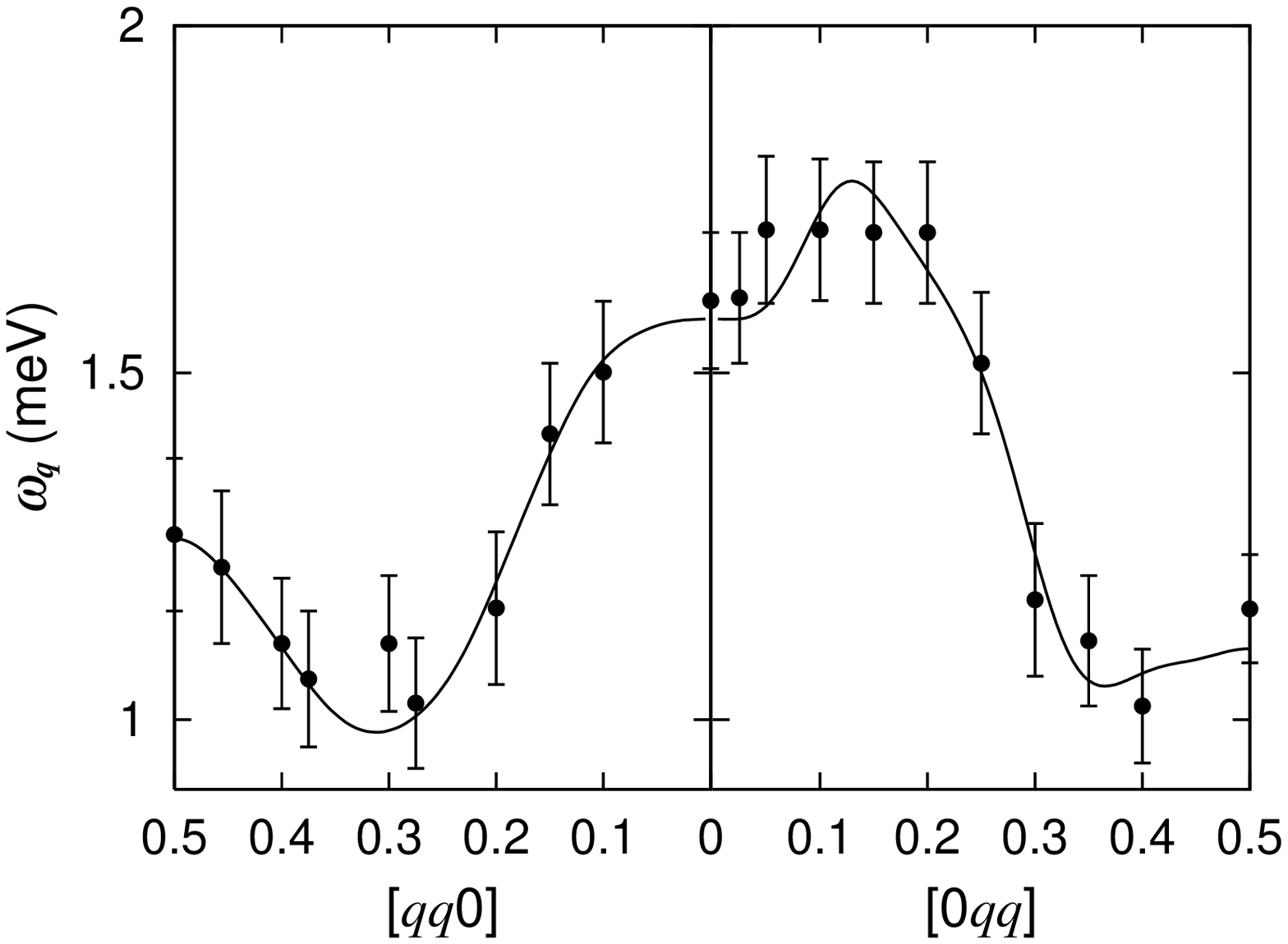}
  \caption{The magnetic excitation spectrum in NdB${}_6$ along
various symmetry directions.
    Experimental data~\cite{erkelens} are shown by solid circles with error bars,
    and the solid lines are theoretical fits. }
\label{fig:fit}
\end{figure}
\begin{figure}
\includegraphics[width=8cm]{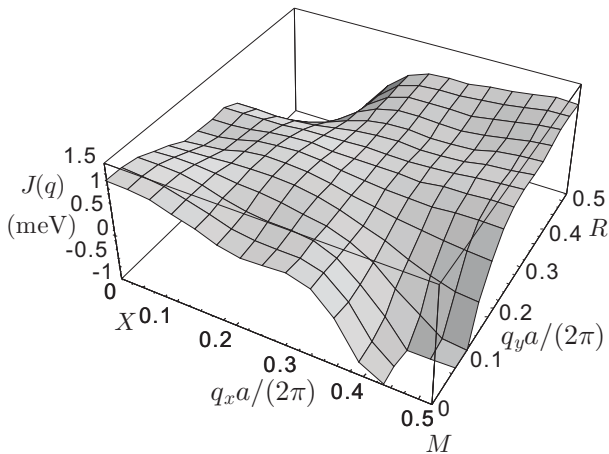}
\caption{Wave-number dependence of $J(\mib q)$
  in the {\it X-M-R} plane. }
  \label{fig:J(q)}
\end{figure}

In view of the fact that the RKKY interaction has a long range, we have included a  sufficient number (= 34) of intersite interactions $J_{ij}$.  The comparison between theory and experiment is shown in Fig.\ref{fig:fit}.
The experimental spectrum is well reproduced by our fit.
The fit gives $J(\mib{Q})=1.16$ meV and $\Delta = 1.47$ meV.
Then we obtain $g_3' = 108$ mK from eq.(\ref{delta}).
This value is in excellent agreement with experimental one, $g_3'\sim$100 mK, deduced from elastic constant \cite{pofahl}.
In the mean-field approximation the N\'{e}el temperature
is not influenced by $g_3'$, and is given by
$$
  T^{\rm MF}_{\rm N}= \frac{4}{3} \frac{\xi^2+\eta^2}{\eta^2}J(\mib{Q}),
$$
which is 9.3 K with the fitted value of $J(\mib{Q})$.

Figure \ref{fig:J(q)} shows $J(\mib{q})$ in the {\it X-M-R} plane.
The maximum of $J(\mib q)$ occurs at $(0,0,1/2)$ in consistency with the A-type order.
In addition, there appears a local maximum near $(1/4,1/4,1/2)$.  The latter indicates a tendency toward the ordering with $\mib q =(1/4,1/4,1/2)$, and
brings about the softening of $\omega_{\mib{q}}$ near this wave number.
We note that the overall behavior of
$J(\mib{q})$, and thus $I(\mib{q}) = (16/3\eta)^2J(\mib q)$, is similar to the result of the interpocket polarization model shown in Fig.\ref{fig:I(q)}.
This similarity supports relevance of the model to real RB$_6$ systems.
The difference should mainly come from our simplification for $W_{\mib {k,p}}$ and $\epsilon_{\mib k}$, and partly from the presence of orbital degeneracy, hybridization, and correlation effect among conduction electrons.

In this paper we have concentrated on the $\mib q$-dependence of the intersite interaction $I(\mib q)$.
As the magnetization grows,
the associated nonlinearity favors a commensurate structure in general.
Then the maximum of $I(\mib q)$ does not necessarily give the ordering wave number at the ground state.
We suggest that the incommensurate-commensurate transition in PrB$_6$ may be interpreted along this line.
It should be worth investigating detailed features which depend on 4f-electron configurations of each rare-earth species.

Another feature to be addressed with finite order parameters is the direction of magnetic moment at each site.  Even with the same $\mib q =(1/4,1/4,1/2)$,
the moment patterns are rather different between CeB$_6$ \cite{effintin} and DyB$_6$ \cite{takahashi}.  While in CeB$_6$
the nearest-neighbor moments are orthogonal to each other and within the (001) plane, the moments in DyB$_6$ point to $(1/2,1/2,1/2)$.  The latter is consistent with the magnetic anisotropy in the paramagnetic region.
In this connection it is interesting to inquire into the spin patterns of GdB$_6$ at low temperature, since the magnetic anisotropy in the paramagnetic region is extremely small \cite{kunii}.

\acknowledgements
This work has been supported partly by Special Coordination Funds for
Promoting Science and Technology, by a Grant-in-Aid for Scientific Research on Priority Areas (B) from MEXT Japan, and by the NEDO international collaboration program ``New boride materials''.



\begin{thebibliography}{130}
\bibitem{effintin}
  J. M. Effantin, J. Rossat-Mignod, P. Burlet,
  H. Bartholin, S. Kunii and T. Kasuya:
  J. Magn. Magn. Mater. {\bf 47-48} (1985) 145.
\bibitem{kusu}
  H. Kusunose and Y. Kuramoto:
  J. Phys. Soc. Jpn. {\bf 70} (2001) 1751.
\bibitem{kuwahara}
  K. Kuwahara {\it et al.}:
  Applied Physics A (to be published).
\bibitem{burlet}
  P. Burlet, J. M. Effantin, J. Rossat-Mignod, S. Kunii
  and T. Kasuya:
  J. Phys. (Paris) C{\bf 8} (1988) 459.
\bibitem{mccarthy}
  C. M. McCarthy and C. W. Tompson:
  J. Phys. Chem. Solids {\bf 41} (1980) 1319.
\bibitem{erkelens}
  W. A. C. Erkelens, L. P. Regnault, J. Rossat-Mignod, M. Gordon,
  S. Kunii, T. Kasuya and C. Vettier:
  J. Phys. (Paris) C{\bf 8} (1988) 457.
\bibitem{schmitt}
  D. Schmitt and P. M. Levy:
  J. Mag. Mag. Mater. {\bf 49} (1985) 15.
\bibitem{onuki}
  Y. Onuki, T. Komatsubara, P. H. P. Reinders and M. Springford:
  J. Phys. Soc. Jpn. {\bf 58} (1989) 3698.
\bibitem{harima}
  H. Harima, O. Sakai, T. Kasuya and A. Yanase:
  Solid State Commun. {\bf 66} (1988) 603, and private communication.
\bibitem{kasuya}
  T. Kasuya and D. H. Lyons:
  J. Phys. Soc. Jpn. {\bf 21} (1965) 287.
\bibitem{teitelbaum}
  H. H. Teitelbaum and P. M. Levy:
  Phys. Rev. B{\bf 14} (1976) 3058.
\bibitem{uimin}
  G. Uimin and W. Brenig:
  Phys. Rev. B{\bf 61} (2000) 60.
\bibitem{pofahl}
  G. Pofahl, E. Zirngiebl, S. Blumenr\"{o}der, H. Brenten,
  G. G\"{u}ntherodt and K. Winzer:
  Z. Phys. B{\bf 66} (1987) 339.
\bibitem{sera}
  M. Sera, S. Itabashi and S. Kunii:
  J. Phys. Soc. Jpn. {\bf 66} (1997) 548.
\bibitem{nakamura}
  S. Nakamura, T. Goto, S. Kunii, K. Iwashita and A. Tamaki:
  J. Phys. Soc. Jpn. {\bf 63} (1994) 623.
\bibitem{lowenhaupt}
  M. Loewenhaupt and M. Prager:
  Z. Phys. B{\bf 62} (1986) 195.
\bibitem{LLW}
  K. R. Lea, M. J. M. Leask and W. P. Wolf:
  J. Phys. Chem. Solids {\bf 23} (1962) 1382.
\bibitem{takahashi}
  K. Takahashi, H. Nojiri, K. Ohoyama,  M. Ohashi, Y. Yamaguchi, S. Kunii and M. Motokawa: Physica {\bf 241-243} (1998) 696;
  K. Takahashi: PhD thesis (Tohoku University, 2002).
\bibitem{kunii}
  R. M. Galera, P. Morin, S. Kunii and T. Kasuya:
  J. Mag. Mag. Mater. {\bf 104-107} (1992) 1336.

\end{thebibliography}
\end{document}